\documentstyle[11pt,epsf,psfig]{article}
\newcommand {\bi}  {\begin{itemize}}
\newcommand {\ei}  {\end{itemize}}
\newcommand {\bc}  {\begin{center}}
\newcommand {\ec}  {\end{center}}
\newcommand {\bq}  {\begin{equation}}
\newcommand {\eq}  {\end{equation}}
\newcommand {\bfi} {\begin{figure}}
\newcommand {\efi} {\end{figure}}
\newcommand {\ba}  {\begin{array}}
\newcommand {\ea}  {\end{array}}
\newcommand {\bt}  {\begin{table}}
\newcommand {\et}  {\end{table}}
\newcommand {\s}     {\ifmmode {\!\!\! /}      \else $\!\!\! /$  \fi}
\newcommand {\sut}   {\ifmmode {\rm SU(2)_L}   \else ${\rm SU(2)_L}$  \fi}
\newcommand {\uo}    {\ifmmode {\rm U(1)_Y}   \else ${\rm U(1)_Y}$   \fi}
\newcommand {\mev}   {\ifmmode {\mathrm MeV}  \else ${\mathrm MeV}$  \fi}
\newcommand {\gev}   {\ifmmode {\mathrm GeV}  \else ${\mathrm GeV}$  \fi}
\newcommand {\tev}   {\ifmmode {\mathrm TeV}  \else ${\mathrm TeV}$  \fi}
\newcommand {\Kz}    {\ifmmode {\rm K^0_s}    \else ${\rm K^0_s}$    \fi}
\newcommand {\Zz}    {\ifmmode {\rm Z^0}      \else ${\rm Z^0 }$     \fi}
\newcommand {\qqbar} {\ifmmode {\rm q\bar{q}} \else ${\rm q\bar{q}}$ \fi}
\newcommand {\ccbar} {\ifmmode {\rm c\bar{c}} \else ${\rm c\bar{c}}$ \fi}
\newcommand {\bbbar} {\ifmmode {\rm b\bar{b}} \else ${\rm b\bar{b}}$ \fi}
\newcommand {\ppbar} {\ifmmode {\rm p\bar{p}} \else ${\rm p\bar{p}}$ \fi}
\newcommand {\uubar} {\ifmmode {\rm u\bar{u}} \else ${\rm u\bar{u}}$ \fi}
\newcommand {\ddbar} {\ifmmode {\rm d\bar{d}} \else ${\rm d\bar{d}}$ \fi}
\newcommand {\ssbar} {\ifmmode {\rm s\bar{s}} \else ${\rm s\bar{s}}$ \fi}
\newcommand {\ttbar} {\ifmmode {\rm t\bar{t}} \else ${\rm t\bar{t}}$ \fi}
\newcommand {\xxbar} {\ifmmode {\rm x\bar{x}} \else ${\rm x\bar{x}}$ \fi}
\newcommand {\rphi}  {\ifmmode {\rm R\phi}    \else ${\rm R\phi}$    \fi}
\newcommand {\ee}    {\ifmmode {e^+e^-}       \else ${e^+e^-}$       \fi}
\newcommand {\mumu}  {\ifmmode {\mu^{+}\mu^{-}}  \else ${\mu^{+}\mu^{-}}$  \fi}
\newcommand {\tata}  {\ifmmode {\tau^{+}\tau^{-}}\else {$\tau^{+}\tau^{-}$}\fi}
\newcommand {\sint}  {\ifmmode {\sin\theta}   \else {\mbox{$\sin\theta$}} \fi}
\newcommand {\cost}  {\ifmmode {\cos\theta}   \else {\mbox{$\cos\theta$}} \fi}
\newcommand {\msbar} {{\ifmmode \overline{MS} \else $\overline{MS}$\fi}}
\newcommand {\lamms} {{\ifmmode \Lambda_{\overline{MS}}
                       \else $\Lambda_{\overline{MS}}$\fi}}
\begin{document}
\title{\huge{Searching for $B_c$ mesons in the ATLAS experiment
at LHC}\thanks{Work partially
supported by CICYT under contract AEN 93-0234 and by IVEI.
}}
\author{{\bf F. Albiol$^{a,b}$\thanks{Now at the Liverpool University,
P.O. Box 147, L693 BX, UK}$\ $, R. P\'erez Ochoa$^{a,b}$\thanks{Now at
CERN, PPE division, CH-1211 Geneva 23}$\ $,
M. A. Sanchis-Lozano$^{a,c}$\thanks{E-mail: mas@evalvx.ific.uv.es}$\ $
and J.A. Valls$^{a,b}$\thanks{Now at Fermilab, P.O. Box 500 Batavia,
Illinois 60510}} \\
\\
\it (a) Instituto de F\'{\i}sica
 Corpuscular (IFIC) Centro Mixto Universidad de Valencia-CSIC \\
\it (b) Departamento de F\'{\i}sica At\'omica, Molecular y Nuclear \\
\it (c) Departamento de F\'{\i}sica Te\'orica \\
\it Dr. Moliner 50, E-46100 Burjassot, Valencia (Spain)}
\maketitle
\abstract{We discuss the feasibility of the observation of the signal
from $B_c$ mesons in the ATLAS experiment of the LHC collider at a luminosity
of ${\approx}\ 10^{33}$cm$^{-2}$s$^{-1}$. In particular we
address the decay mode $B_c{\rightarrow}J/\psi \pi$ followed by the leptonic
decay $J/\psi{\rightarrow}\mu^+\mu^-$, which should permit an accurate
measurement of the $B_c$ mass. We performed a Monte Carlo study of the signal
and background concluding that a precision of $40$ MeV for the $B_c$ mass
could be achieved after one year of running.}
\large
\vspace{-11cm}
\begin{flushright}
{\bf IFIC/95-24 }
\end{flushright}
\vspace{11cm}
\section{Introduction}
There is a general consensus in the scientific community \cite{aach} that
the scope of a future high-luminosity, high-energy hadron collider
like LHC should not be restricted to the hunting of the standard model
Higgs and its extensions, or the search for supersymmetry. Other topics
requiring lower luminosities like top and beauty physics deserve in their
own right a close \vspace{0.1in} attention.\par
In this paper we shall not dwell on general aspects of $B$ physics already
treated in detail in references \cite{ee}, but rather we
shall focus more specifically on the observation of $B_c$ ($\overline{b}c$)
mesons. With regard to a general description of the ATLAS detector,
technical details and foreseen performances we refer the reader to
the Technical Proposal of the ATLAS Collaboration \cite{tp} and
references therein. Finally let us observe that the aim of this paper
is to extend and update our previous preliminary work on the detection of
$B_c$ mesons presented as an ATLAS internal note \vspace{0.1in} \cite{note}.
\par
 From the theoretical point of view, $B_c$ mesons exhibit some unique features
making them especially suitable for the study of the strong interaction
dynamics in hadrons. First, $B_c$ states occupy an interpolating position in
hadronic spectroscopy between charmonium and bottomonium resonances
\cite{bagan}. QCD-inspired theories like potential models can be submitted to
a close scrutiny in such intermediate mass region, besides with different
constituent quark masses. Moreover, low-energy effective theories like
nonrelativistic QCD, which can be formulated on the lattice, may be applied
as well either to heavy quarkonium formation or decay \vspace{0.16cm}
\cite{bodwin}.
\par
Furthermore, in contrast to singly heavy hadrons ($D,\ B,\
{\Lambda}_c,\ {\Lambda}_b,\ ...$) both constituent quarks can undergo a
weak decay, permitting a test of the $\lq\lq$spectator" behaviour. (Let
us note, nevertheless, that a simple-minded spectator model would not be
valid since both heavy masses are involved in the hadron dynamics, even as an
asymptotic \vspace{0.01in} limit.)

\section{$B_c$ signal}
At the center-of-mass energy $\sqrt{s}=14$ TeV, the cross-section for
beauty production is assumed to be $500$ $\mu$b leading to
$5{\times}10^{12}$ $b\overline{b}$ pairs per year-run ($10^7$ s) at a
luminosity of $\cal{L}\ {\approx}\ 10^{33}$ cm$^{-2}$ s$^{-1}$,
corresponding to an integrated luminosity of ${\sim}\ 10$ fb$^{-1}$. The
number of bottom pairs reduces, however, to $2.3{\times}10^{10}$
by requiring events with a triggering muon coming from either a $b$ or a
$\overline{b}$ under the kinematic cuts \footnote{The transverse
momentum $p_{\bot}$ is measured with respect to the beamline and the
pseudorapidity is defined as ${\eta}=-\ln{\tan{\theta/2}}$ where $\theta$
is the polar angle} $p_{\bot}>6$ GeV/c and ${\mid}{\eta}{\mid}<2.2$
\vspace{0.1in} \cite{tp}.
\par
On the other hand, assuming that the $b$-quark fragmentation yields a $B_c$
or a $B_c^{\ast}$ with probability of the order of $10^{-3}$ \cite{pro}, the
yield of $B_c$ mesons (not yet triggered) per year of running would be
roughly ${\simeq}\ 10^{10}$.
\vspace{0.3cm}
\subsection*{$B_c\ {\rightarrow}\ J/\psi\ \pi$ channel}
This exclusive channel followed by the leptonic decay of the
$J/\psi$ resonance into a pair of oppositely charged muons offers
several important advantages. First of all, it allows for the mass
reconstruction of the $B_c$ meson. Observe also that anyone of the two muons
can trigger the decay. Besides, it is very clean topologically with a common
secondary vertex for all three charged particles, two of them (the muons) with
the additional constraint of their invariant mass compatible with a
$J/\psi$. Furthermore, the expected branching fraction is not too small, about
$0.2\%$ \cite{lusi}, which combined with the branching
fraction of the leptonic decay of the resonance
$BR(J/\psi\ {\rightarrow}\ {\mu}^+{\mu}^-)\ {\simeq}\ 6\%$,
yields an overall branching fraction for the signal of $10^{-4}$. Thus, the
number of such events turns out to be ${\simeq}\ 10^6$ per year of
\vspace{0.3cm} running.

\subsection*{$B_c\ {\rightarrow}\ J/\psi\ \mu^+ \ {\nu}_{\mu}$ channel}
In spite of the fact that this channel does not permit
the measurement of the $B_c$ mass, its signature would be quite clean
experimentally when the $J/\psi$ decays into a pair of muons, providing
a three muon vertex. The overall branching fraction is then
${\simeq}\ 10^{-3}$ \cite{masl}. We leave the analysis and
physical interest of this decay
mode to a separate publication \cite{maga}.

\section{Detection efficiencies and background}
A study has been performed in order to estimate the signal
detection efficiency and background for the
$B_c\ {\rightarrow}\ J/\psi({\rightarrow}\ {\mu}^+{\mu}^-)\ \pi$ channel. The
Monte Carlo employed for the signal simulation corresponds to a sample of
$B_c\ {\rightarrow}\ J/\psi\pi$ events \footnote{We are here interested
in the efficiency of the signal rather than in absolute production rates for
which we made a rough estimate in the previous section according to
fragmentation of $b$ quarks into $B_c$ mesons} while for the background we
have used a sample of inclusive $b$ muon decays generated in all the cases
with PYTHIA 5.7 \vspace{0.16cm}.
\par
A full simulation of the ATLAS detector using the GEANT Monte Carlo
program was performed to study in detail the reconstruction of $B$ events,
parametrizing the detector effects through smearing routines
\cite{comu}. Consequently, our signal and background analysis were performed
using a particle-level simulation with parametrized momentum and impact
parameter resolutions \cite{tp}.
\vspace{0.3cm}
\par
In this paper, we shall consider two types of \vspace{0.16cm}background:
\footnote{Muons coming from semileptonic decays of long-lived
particles such as pions or kaons contribute in a negligible amount to
trigger rates. On the other hand $B$ decays into $J/\psi$ and a charged
particle would give an invariant mass quite below the $B_c$ mass so they
are of no concern}
\begin{itemize}
\item[a)] Combinatorial background due to muons from semileptonic decays of
$b\overline{b}$ pairs produced at the main interaction. Cascade
contributions such as $b{\rightarrow}c{\rightarrow}\mu$ are included as
well for random combinations with any other muon in the same event.
\item[b)] Contamination from prompt $J/\psi$'s
in combination with another charged hadron (interpreted as a pion)
from the main vertex. (In fact data released by Tevatron on the $J/\psi$
yield point out a production rate quite larger than initially expected
\cite{greco}.) Incorrect tracking may give rise to the
reconstruction of a (fake) secondary vertex, becoming a potential
source of a large amount of \vspace{0.16cm} background.
\end{itemize}
\par
In a first step, we imposed the following cuts on events based on
kinematic constraints:
\begin{itemize}
\item $p_{{\bot}min}(trig.\mu)=6$ GeV/c \, ;
\, $\mid\,{\eta}_{max}(trig.\mu)\mid
=2.2$
\item $p_{{\bot}min}(\mu)=3$ GeV/c \, ; \, $\mid\,{\eta}_{max}(\mu)\,\mid =
2.5$
\item $p_{{\bot}min}(\pi)=1$ GeV/c \, ; \, $\mid\,{\eta}_{max}(\pi)\,\mid =
2.5$
\item $M_{{\mu}^+{\mu}^-}=M_{J/\psi}\ {\pm}\ 50$ MeV
\end{itemize}
\vspace{0.21cm}
The two first cuts correspond to the requirement of the 1st-level $B$ physics
trigger leading in our case to an efficiency of $\sim 15\%$ in triggering
one of the two muons from the $J/\psi$. We next take into account the
detection efficiency
for the signal after applying the rest of $p_{\bot}$ and $\eta$ cuts which
turns
out to be $\sim 21\%$. Setting the efficiency for muon identification as
$80\%$ and the track reconstruction as $95\%$ \cite{tp} we get a
combined detection efficiency of $\sim 2\%$ leading to an observable signal of
about $20,000$ events per \vspace{0.16cm} year of running.
\par
The last of the cuts described above constrains the two muons invariant
mass to be compatible (within two standard deviations \cite{tp})
with the nominal $J/\psi$ mass, thus drastically
reducing random combinations. However, background of class $b)$ can
potentially pass all the kinematic cuts by a large amount, so another type
of rejection is \vspace{0.1in} required.
\par
To this end, we adapted to our needs the vertex reconstruction (i.e.
vertex finding and fitting) routines of the LEP experiment DELPHI at
CERN \cite{bil} \footnote{We are indebted to E.
Cortina for technical advice in this task}. The vertex fitting algorithm
provides as output the coordinates of the secondary vertex, the track momenta
re-evaluated with the vertex constraint and the goodness of the fit
by means of the total ${\chi}^2$ as well as the contribution of each
single track to it. In particular, we employed for background rejection
the three spatial coordinates and the ${\chi}^2$ per degree of freedom
for each fitted secondary
vertex formed by the two muons and the charged hadron (assumed to be a pion)
satisfying the above kinematic constraints. The distance
between the reconstructed vertex and the primary ($pp$) interaction point
was thereby determined. We shall refer to it as the decay
length even for background events of \vspace{0.1in} class b).
\par
Hence, candidate (either signal or background) events were
required to pass the following extra \vspace{0.16cm} cuts:
\begin{itemize}
\item total ${\chi}^2\ <\ {\chi}_0^2$
\item ${\chi}_i^2\ <\  \frac{{\chi}_0^2}{3}$ for each single track-$i$
\item decay length larger than $L_0$
\end{itemize}
\vspace{0.2cm}
where ${\chi}_0^2$, $L_0$ have to be optimized to remove the background
as much as possible but with a good acceptance for the signal. In our
analysis we found ${\chi}_0^2=8$ and \vspace{0.1in} $L_0=350$ $\mu$m.
\par
In figure 1 the effects of these cuts on signal and
background events are shown. Let us remark that with this selected range of
${\chi}^2$ and decay length, the background of class $a)$ falls to
the $4.5\%$ (with respect to the background after the kinematic cuts), the
background of class $b)$ is completely removed, whereas the acceptance
for the signal turns out to be $46\%$. The total rejection
\footnote{No candidate event was found in a sample corresponding to
$100,000$ prompt $J/\psi$'s} of the contamination from
prompt $J/\psi$'s can be quickly understood since actually they are produced
(thus decaying) at the $pp$ interaction vertex itself. Those fake
secondary vertices $\lq\lq$reconstructed faraway" from the main vertex
give rise to a quite large ${\chi}^2$ in the fit, so being removed by these
combined \vspace{0.1in} cuts.
\par

Figure 2 shows the reconstructed $(\mu^+\mu^-)_{J/\psi}\ \pi$
mass distribution for the expected signal above the surviving background
once all the cuts have been applied, for an
integrated luminosity of \vspace{0.4cm} $10$ fb$^{-1}$.
\par

\section{Summary}
We have found that the self-triggering weak decay
$B_c{\rightarrow}J/\psi\ \pi$, followed by the leptonic decay of the $J/\psi$
into two muons, could be clearly observed in the ATLAS detector at LHC. Under
rather conservative assumptions, a total number of ${\approx}\ 10,000$
signal events could be fully reconstructed after one year run,
corresponding to $10$ fb$^{-1}$ at $\lq\lq$low" luminosity
(${\approx}\ 10^{33}$ cm$^{-2}$ s$^{-1}$). This represents a signal to
background ratio of about $0.5$  with a statistical significance of
${\approx}\ 20$ standard deviations above a nearby almost flat
background. The foreseen mass resolution of the $B_c$ meson is about $40$ MeV.
\vspace{0.16cm}
\par

\vspace{0.7cm}
\section*{Acknowledgments}
We are indebted to P. Eerola and N. Ellis and the B-Physics Group of
the ATLAS Collaboration for their interest and advice. One
of us (M.A.S.L.) thanks the Vice-Rectorado
de Investigaci\'on de la Universitat de Val\`encia for partial
\vspace{0.7cm} support.

\newpage

\bfi[hbt]
\centerline{\psfig{figure=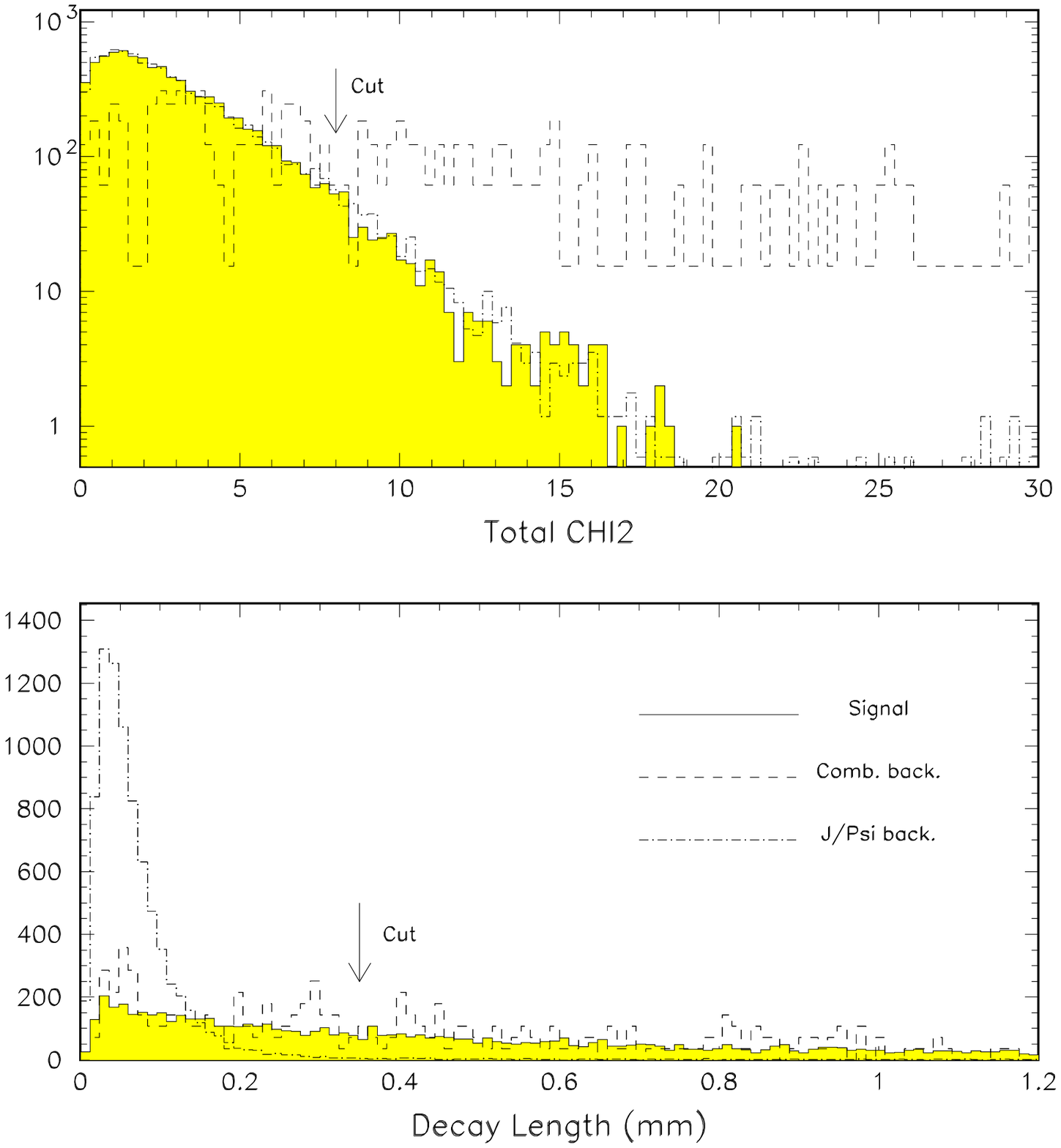,height=15.5cm,width=15.5cm}}
\caption{Effect of the vertex reconstruction cuts on the signal (solid
line) and background of class $a)$ (dashed line) and class $b)$ (dot-dashed
line), separately. The cut on the ${\chi}^2$ was set
equal to 8 and the decay length cut was $350$ $\mu$m. All three samples are
normalized to the same number of events.}
\label{fig:bcsign}
\efi

\newpage

\bfi[hbt]
\centerline{\psfig{figure=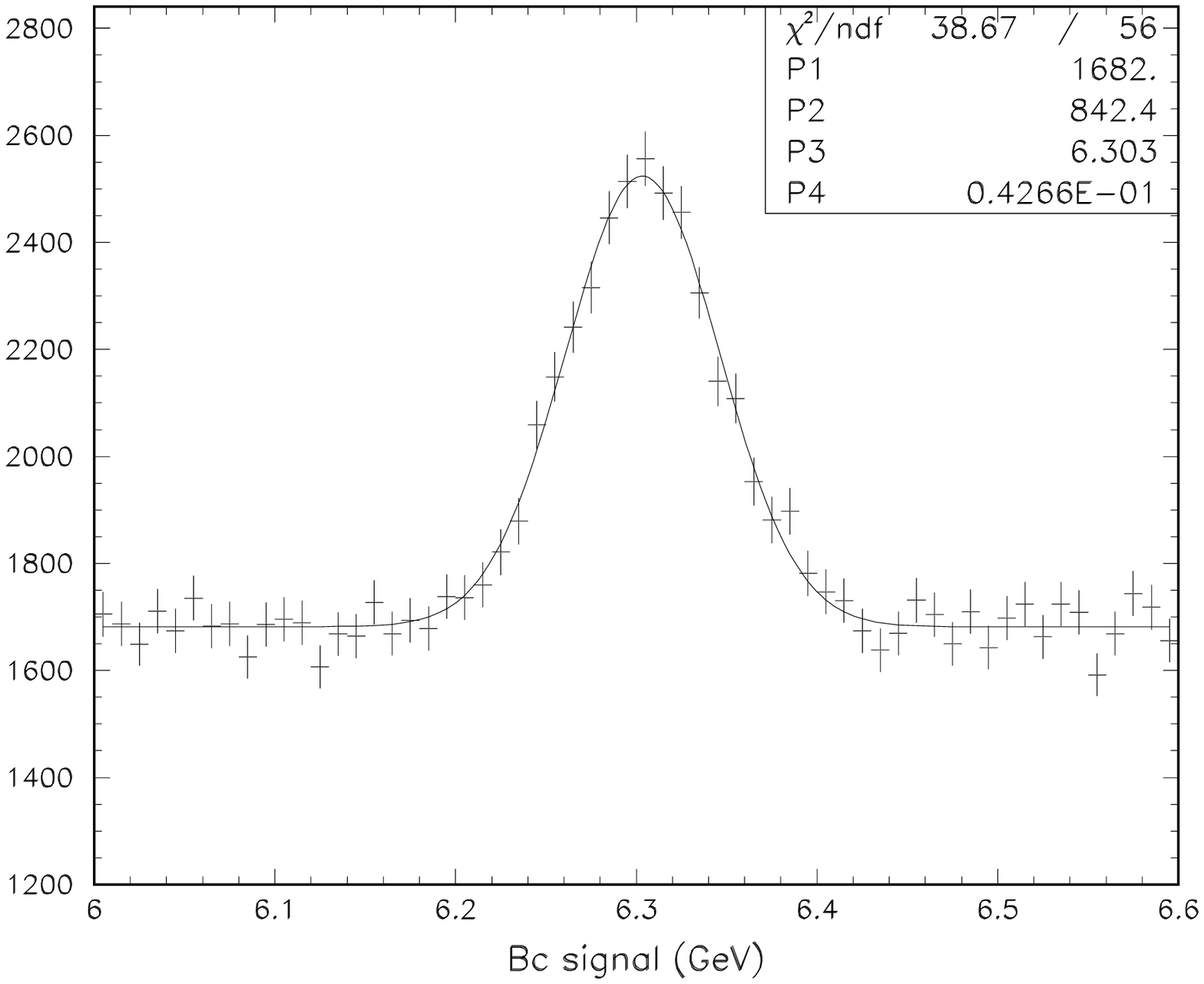,height=15.5cm,width=15.5cm}}
\caption{Reconstructed $({\mu}^+{\mu}^-)_{J/\psi}\ \pi$ mass
distribution after cuts. The nominal value for the $B_c$ was
set equal to 6.3 GeV.  The solid line corresponds to a linear+Gaussian fit.}
\efi

\end{document}